\documentclass[preprint2]{aastex}

\slugcomment{Not to appear in Nonlearned J., 45.}

\shorttitle{Collapsed Cores in Globular Clusters}
\shortauthors{Djorgovski et al.}

\begin{document}


\title{A Self-Consistent Explanation of TeV Emissions from HESS J1640-465 and HESS J1641-463}


\author{Yunyong Tang}
\affil{Department of Physical Science and Technology, Kunming University, Kunming 650214, China}
\author{Chuyuan Yang}
\affil{Yunnan Observatories, Chinese Academy of Sciences, Kunming 650011, China
\\Key Laboratory for the Structure and Evolution of Celestial Objects, Chinese Academy of Sciences}
\author{Li Zhang}
\affil{Department of Astronomy, Key Laboratory of Astroparticle Physics of Yunnan Province, Yunnan University, Kunming, 650091, China\email{lizhang@ynu.edu.cn}}
\author{Jiancheng Wang}
\affil{Yunnan Observatories, Chinese Academy of Sciences, Kunming 650011, China
\\Key Laboratory for the Structure and Evolution of Celestial Objects, Chinese Academy of Sciences}



\begin{abstract}
The bright TeV source HESS J1640-465 is positionally coincident with the young SNR G338.3-0.0, and the nearby HESS J1641-463 with TeV gamma-ray emission seems
to be closely associated with it. Based on the nonlinear diffusion shock acceleration (NLDSA) model, we explore the emission from these two TeV sources,
the particle diffusion is assumed to be different inside and outside the absorbing boundary of the particles accelerated in the SNR shock. The results
indicate that (1) the GeV to TeV emission from the region of the HESS J1640-465 is produced as a result of the particle acceleration inside the SNR G338.3-0.0;
and (2) the runaway cosmic-ray particles outside the SNR are interacting with nearby dense molecular cloud (MC) at the region of the HESS J1641-463, corresponding $\pi^0$ decay gamma-ray in proton-proton collision contribute to the TeV emission from the HESS J1641-463. Also we investigate the possible X-ray emission in molecular cloud from synchrotron procedure by secondary $e^\pm$  produced through escaped protons interaction with the MC.
\end{abstract}





\section{Introduction}           

It is generally believed that the particles in supernova remnants (SNRs) can be efficiently accelerated to relativistic energy  through the diffusive shock
acceleration (DSA) mechanism. Although there is no direct evidence to show that the maximum energy of accelerated particles in SNRs can reach a so called knee energy $\sim10^{15}$ eV \citep[e.g.][]{Blasi13}, {\it Fermi} observations show that the gamma-ray spectra of some SNRs (e.g., IC 443 and W44) appear the characteristic pion-decay feature, resulting in  direct evidence that cosmic-ray protons are accelerated in SNRs \citep{A13}. Therefore, the observations of high energy $\gamma$-ray emissions from SNRs will help us to understand the particle acceleration mechanisms in SNRs.

Recently, several TeV gamma-ray sources (possibly) associated with SNRs in the Galaxy have been detected by  the High Energy Stereoscopic System (H.E.S.S.) \citep[e.g.][]{Abramowski2014a,Abramowski2014b,Abramowski2015a,Abramowski2015b}, which indicate that a SNR with TeV emission can accelerate particles up to higher energy compared with a SNR without TeV emission (e.g., Puppis A \citep{Abramowski2015c}) if the TeV emission comes from the $\pi^0$ decay via proton-proton collision. Here we will focus on the origins of TeV gamma-rays from two TeV gamma-ray sources  HESS J1640-465 and HESS J1641-463, the latter is very close to the former and is surrounded by a dense molecular cloud (MC).

HESS J1640-465 is located at the ambient region of a younger (1.1-3.5 kyr) SNR G338.3-0.0 \citep[e.g.][]{Shaver1970,Whiteoak1996},
it is an extended source with very high-energy gamma-ray emission discovered by the H.E.S.S during its survey of the Galactic Plane in 2004-2006 (Aharonian et al. 2006).
Recent study indicate that very high energy (VHE) gamma-ray emission from HESS J1640-465 sufficiently overlaps with the north-western part of the SNR shell of G338.3-0.0, moreover this VHE gamma-ray spectrum naturally connects with the GeV spectrum and has a high energy cut-off \citep{Abramowski2014a}. Combined with the observed data of this source by the XMM-Newton, \citet{Abramowski2014a} pointed out that the particles can be accelerated up to tens of TeV energies in emission region, and the morphology of GeV-TeV
emission imply some possible evidence for proton acceleration in the SNR shell of G338.3-0.0. HESS J1641-463 is positionally coincident with a older (5-17 kyr) SNR G338.5+0.1 and appears a hard spectrum in VHE domain \citep{Abramowski2014b}. Because of its low brightness and proximity to HESS J1640-465, the observed feature of HESS J1641-463 is
confusing, and  the obvious X-ray candidate is not found as an evident association so far. Therefore, \citet{Abramowski2014b} proposed three possible scenarios to explain the potential origin of VHE hard spectrum of HESS J1641-463 based on a simple analytic method. Here we will consider the second scenario proposed by \citet{Abramowski2014b}, i.e.,
the adjacent younger SNR G338.3-0.0 would be a major source of cosmic rays since SNR G338.5+0.1 is older (5-17 kyr), the escaping protons
from the SNR G338.3-0.0 interact with the dense MC in the region of the HESS J1640-465, subsequently the high energy gamma-rays are produced through $\pi^0$ decay in proton-proton collision.

In this scenario, we need to know not only the particle spectra accelerated inside the SNR but also that escaping from the SNR. For the particle acceleration in the SNR, non-linear diffusive shock acceleration (NLDSA) models have been widely used \citep[e.g.][]{Kang2006,Zirakashvili2010,Caprioli2008,Telezhinsky2012a,Ferrand2014}, which also involved the escape of high-energy particles \citep[e.g.][]{Ellison2011,Telezhinsky2012b,Kang2013}. Recently, based on the NLDSA model of \citet{Zirakashvili2012} and the assumption that the particle diffusion is different inside and outside the absorbing boundary of the particles accelerated in the SNR shock, \citet{Yang2015} investigated the escape and propagation of high-energy protons near young SNRs. In this paper, we will self-consistently explain the high energy gamma-ray emissions from these two TeV sources in the frame of the model given by \citet{Yang2015}. In Section 2, we review briefly the NLDSA model used here. We show the model application and the corresponding results in Section 3.
Finally we give our conclusion and discussion in Section 4.

\section{The Model}
\label{sect:Mod}

\subsection{Particle Spectrum Accelerated in a SNR}

\citet{Zirakashvili2012} described a numerical model of the NLDSA in a SNR. In this model, hydrodynamic equations for gas denisity $\rho(r, t)$, gas velocity $u(r, t)$, and gas pressure $P_{\rm g}(r, t)$ (see Eqs. (1) -(3) of their paper), combined with the equation for the quasi-isotropic particle momentum distribution $N(r, t , p)$ presenting the transport and acceleration of relativistic particles by the forward shock and reverse shocks (see Eq. (4) of their paper), are solved numerically in the spherically symmetric case.
Since the pressure gradient $\partial P_{\rm c}/\partial r$ ($P_{\rm c} = 4\pi\int p^2dp v pN(r, t , p)/3$) of accelerated particles with a momentum $p$ and a velocity $v$ is included in the hydrodynamic equations (see Eqs. (2) -(3) of \citet{Zirakashvili2012}), the accelerated particles will react onto the shock structure, resulting in a self-regulation of acceleration efficiency. \citet{Zirakashvili2012} described the numerical procedure of solving above equations in detail.

We now present the relevant physical parameters in this model. Firstly, thermal protons with the momenta $p = p_f$ and $p = p_b$ are assumed to be injected at the fronts of the forward and reverse shocks at $r = R_f(t)$ and $r = R_b(t)$, respectively, where the dimensionless parameters $\eta_f$ and $\eta_b$ are corresponding injection efficiencies and subscript $f$ and $b$ for quantities are represented the forward and reverse (backward) shock respectively. Secondly, the accelerated protons propagate inside the SNR through advection and diffusion. On the one hand, the advective velocity of the protons is $w(r, t) = u(r, t) + \xi_{A}V_A/\sqrt{3}$ in the isotropic random magnetic field $B$, where $V_A = B/\sqrt{4\pi\rho}$ is {\bf Alfv\'en} velocity and $\xi_A$ is a parameter which represents the relative direction of fluid velocity to {\bf Alfv\'en} velocity with
$\xi_A = -1$ in upstream of forward shock, $\xi_A = 1$ in upstream of reverse shock and $\xi_A=0$ in the other region (e.g. the downstream for forward and reverse
shocks ). On the other hand, the diffusion coefficient of protons inside the SNR can be expressed as $D_{ f,b}=\eta _BD_B(1+\frac{p}{P_{ f,b}})$, where $\eta_B = 2$, $D_B=vpc/3qB$ is Bohm diffusion coefficient, $P_f$ and $P_b$ are the momenta separating proton diffusion in forward and reverse shock regions (the detailed description see \citet{Zirakashvili2014}). Below $P_f=0.61$~TeV/$c$ and $P_b=6.1$~TeV/$c$ are assumed to reproduce the gamma ray observations.

Although the modeling of amplification and transport of magnetic field cannot be taken into account for consistency, the magnetic field has been introduced to play dynamical role in downstream regions both reverse and forward shock, the pressure and energy flux of magnetic field cannot be ignored due to strong magnetic field, because they have effects on the dynamical evolution of the SNRs shock, therefore the magnetic pressure and magnetic energy flux should be introduced to the fluid dynamical equation \citep{Zirakashvili2014} in downstream areas, and the magnetic field spatial distribution depending on radius at $r > R_f$ is
\begin{equation}
B(r)=\sqrt{4\pi \rho _0}\frac {V_f}{M_{A,f}}\left(\frac {\rho(r)}{\rho _0}\right)^{\gamma_m/2}\;.
\label{mag_eq}
\end{equation}
Here $\rho _0$ is the gas density of the circumstellar medium. The parameter $M_{A,f}$  determines the value of the amplified magnetic field strength at forward shock, in the shock transition region the magnetic field strength is derived via Hugoniot condition, i.e. in downstream region the magnetic field  strength is increased by a factor of $\sigma^{\gamma_m/2}$ , where $\sigma$ is the shock compression ratio. In the upstream region  of the reverse shock we also used the equation similar to Eq. (\ref{mag_eq}), only replaced $V_f$ and $M_{A,f}$
with $V_b$ and $M_{A,b}$. In following application, we use the adiabatic index of isotropic random magnetic field $\gamma_m=4/3$, the strength factors of magnetic field $M_{A,f}=4$ and $M_{A,b}=8$.

\subsection{Particle Spectrum Escaped from a SNR}

The NLDSA with the reaction of accelerated particles onto the shock
structure has been proposed in detail to research the radiation process in SNRs \citep{Zirakashvili2012}.
Subsequently, the processes of accelerating particles in shock front and
the escape of high energy particles at a absorbing boundary have been taken into account \citep[e.g.][]{Telezhinsky2012b,Yang2015}.

To self-consistently explain VHE emissions from HESS J1640-465 and HESS J1641-463, we deal with the particle diffusion by dividing the simulating domains into two diffusion regions \citep[also see the first model of][]{Yang2015}. one is Bohm-like type in the vicinity of the forward shock within a radius of $2R_f$ (inside SNRs), the other is generally considered as Galactic diffusion when accelerated particles is far away the Bohm limited region (e.g. $r >2R_f$ is outside SNRs). The propagation of high energy particles outside a SNR is considered as Galactic diffusion, and the diffusion coefficient can be expressed as
\begin{equation}
D_g(E_p) = 10^{28}~\chi\left(\frac{E_p}{10~\rm{GeV}}\right)^{\delta}~\rm{cm}^2 \rm{s}^{-1}\;,
\end{equation}
where $E_p$ is the accelerated proton energy, {\bf $\chi$ is the correction
factor in the case of slow diffusion around the SNRs, its value ($\chi<1$) is lower than that of average Galatical diffusion case ($\chi\sim 1$) (see \citet{Gabici2007}, also see their arguments of \citet{LiChen2012} and \citet{Ohira2011}), and $\delta \approx 0.3-0.7$ is  the energy-dependent index of the diffusion coefficient
\citep[e.g.][]{Berezinskii1990,Fujita2009}. Below we use the value of $\delta=0.5$ and $\chi=0.1$.}

\subsection{Intrinsic $\gamma$-ray emission inside a SNR}

With the NLDSA model, the hydrodynamic process and particle propagation can be calculated
by means of combining the evolution of magnetic field and particle diffusion. Thus, for a given SNR, the spectra of accelerated particles are produced naturally at any time $t$ . In general, high energy gamma-rays are thought to originate from two scenarios, one is hadronic scenario, the radiation are produced through the proton-proton interaction process, and the other is inverse-Compton (IC) process by energetic electrons scattering up soft photons, this also called leptonic scenario. In this paper, the seed photons of IC come
from 2.7 K microwave background light, and we will use the emissivity formula developed by \citet{Blumenthal1970}, the proton-proton interaction process which can be calculated by the analytic approximation given by \citet{Kelner2006}.

\subsection{Nonthermal photons in a MC nearby a SNR}

The escaped protons outside a SNR during their propagations in Galaxy are assumed to collide a nearby MC, so the proton-proton interaction becomes an important process producing nonthermal photons in the MC. Such a process produce neutral pions ($\pi^0$) and charged pions ($\pi^\pm$). On the one hand, high energy gamma-rays can be created through $\pi^0$ decay \citep[e.g.][]{Kelner2006,ZF07,TFZ11}. On the other hand, $\pi^\pm$ decay will produce secondary electrons and positrons, and these $e^\pm$ pairs will emit nonthermal photons through synchrotron radiation and inverse Compton scattering. Generally a molecular cloud has a density $n_{\rm MC} > 10^2~\rm{cm}^{-3}$ and magnetic field $B_{\rm MC} > 10^{-5}$G, {\bf if the produced $e^\pm$ pairs diffuse in the MC with a radius $R_{\rm MC}$, then the typical diffusive time-scale is $t_{\rm dif} \sim R_{\rm MC}^2/(6D_g(E))\sim 3\times 10^3(R_{\rm MC}/10~\rm pc)^2\chi^{-1}(E/10~\rm GeV)^{-\delta} (B_{\rm MC}/100~\mu{\rm G})^{0.5}$ yr. At the same time, the energy loss time for electron synchrotron emission is given by $t_{\rm syn}=1.3\times10^3(E/{\rm TeV})^{-1}(B_{\rm MC} /100~\mu{\rm G})^{-2}$ yr \citep[see][]{Gabici2009}. With typical parameters $R_{\rm MC}\sim10$ pc, $\chi=0.1$, and $\delta=0.5$, the ratio of $t_{\rm dif}$ to $t_{\rm syn}$ is approximately $t_{\rm dif}/t_{\rm syn}\approx 2(E/{\rm TeV})^{0.5}(B_{\rm MC}/100~\mu {\rm G})^{2.5}$. Obviously, for strong magnetic field ($>100~\mu {\rm G}$), the synchrotron loss time for electrons with their energies $\sim {\rm TeV}$ is shorter than the diffusion time, the secondary electrons would radiate all their energy through synchrotron emission, thus the diffusive transport of electrons within the MC can be neglected.

The density of the MC should satisfy some distributions \citep{Protheroe2008}, the density profiles could affect the distribution of magnetic field, particle diffusion and the following non-thermal emissions. For simplicity, in this paper, we assume a flat density profile of the MC with an average value of $n_{\rm MC} = 10^2~\rm{cm}^{-3}$.
}

After obtaining the production rate, $q(E)_\pm$, of secondary $e^\pm$, we can readily obtain the number densities of electrons and positrons per unit energy \citep{Jones2008}, $N^{\pm}(E)$, within the molecular cloud:
\begin{equation}
N_{\pm}(E)=\frac{\int_E^\infty q_\pm (E^\prime)dE^\prime}{dE/dt}\;,
\end{equation}
where $dE/dt$ is the total energy loss rate of electrons at given energy $E$ due to synchrotron emission.
\begin{equation}
dE/dt=\frac{4}{3}\sigma_\tau c\beta U_B \gamma^2,
\end{equation}
where $U_B=B_{\rm MC}^2/8\pi$ is the density of magnetic field within the molecular cloud. Then the synchrotron emissivity $j_\nu$ is calculated by using formulae in synchrotron radiation theory \citep{Rybicki1979}. At the same time, the secondary $e^\pm$ also generated IC gamma-ray emission, the process and seed photons is the same as the corresponding description of Sect. 2.3.

\section{Application}
\label{sect:App}

In this section, we apply the model to explore the high energy emission origin of
the two TeV sources: HESS J1640-465 and HESS J1641-463. It is thought that the particles
are accelerated inside SNR 338.3-0.0, then high energy gamma-rays yielded
as a result of $\pi^0$ decay in the proton-proton interaction process in SNR,
these gamma-rays could just present at the region of HESS J1640-465.
In addition, some high energy particles can escape out the SNR G338.3-0.0 at the region of $r > 2R_f$,
the runaway particles will collide with the mater in dense MC by
proton-proton interaction (also see the discussion of \citet{Abramowski2014b},
and then some high energy gamma-ray could be produced at the area of HESS J1641-463.

\begin{figure}
\centering
\includegraphics[width=9.0cm, angle=0]{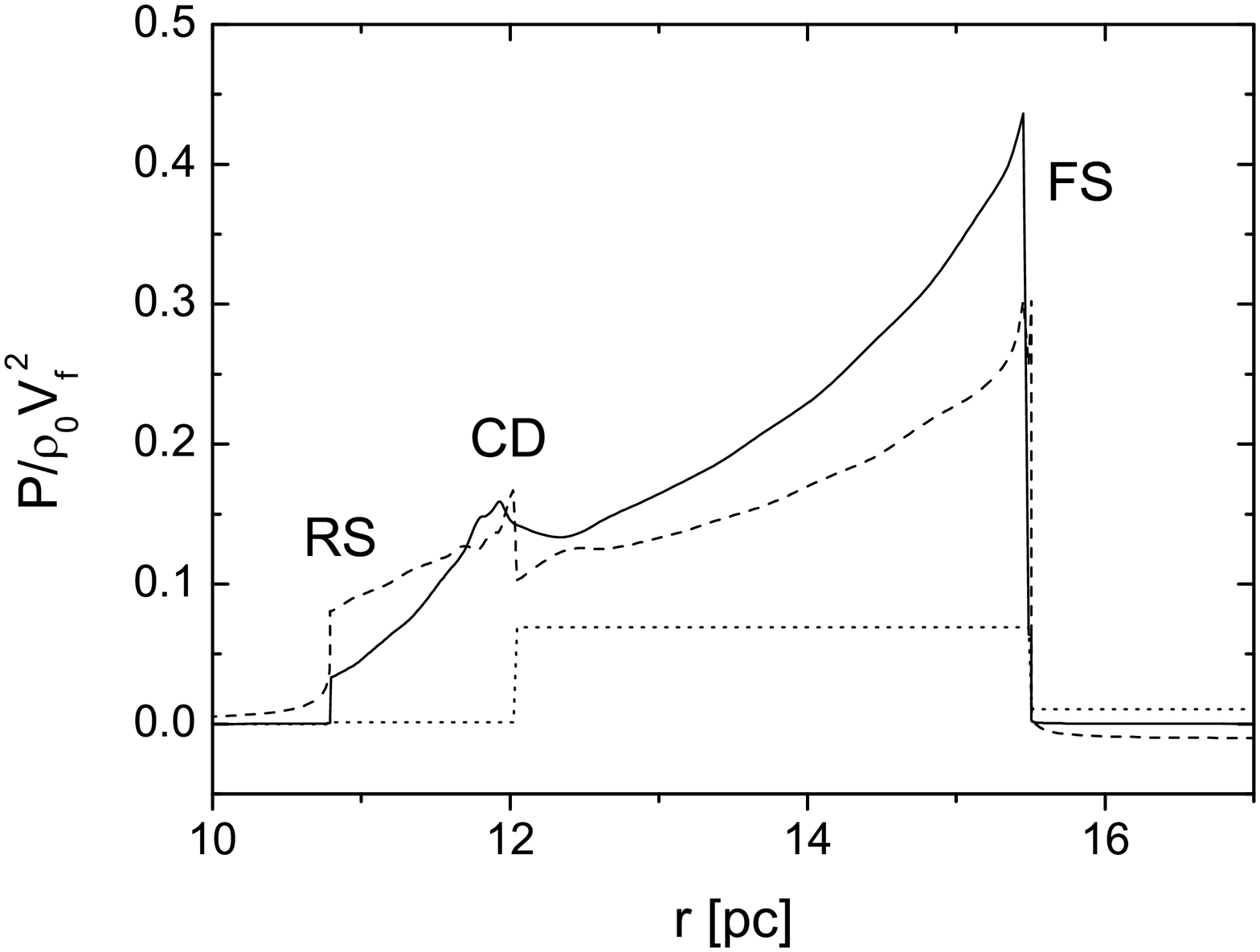}
\includegraphics[width=9.0cm, angle=0]{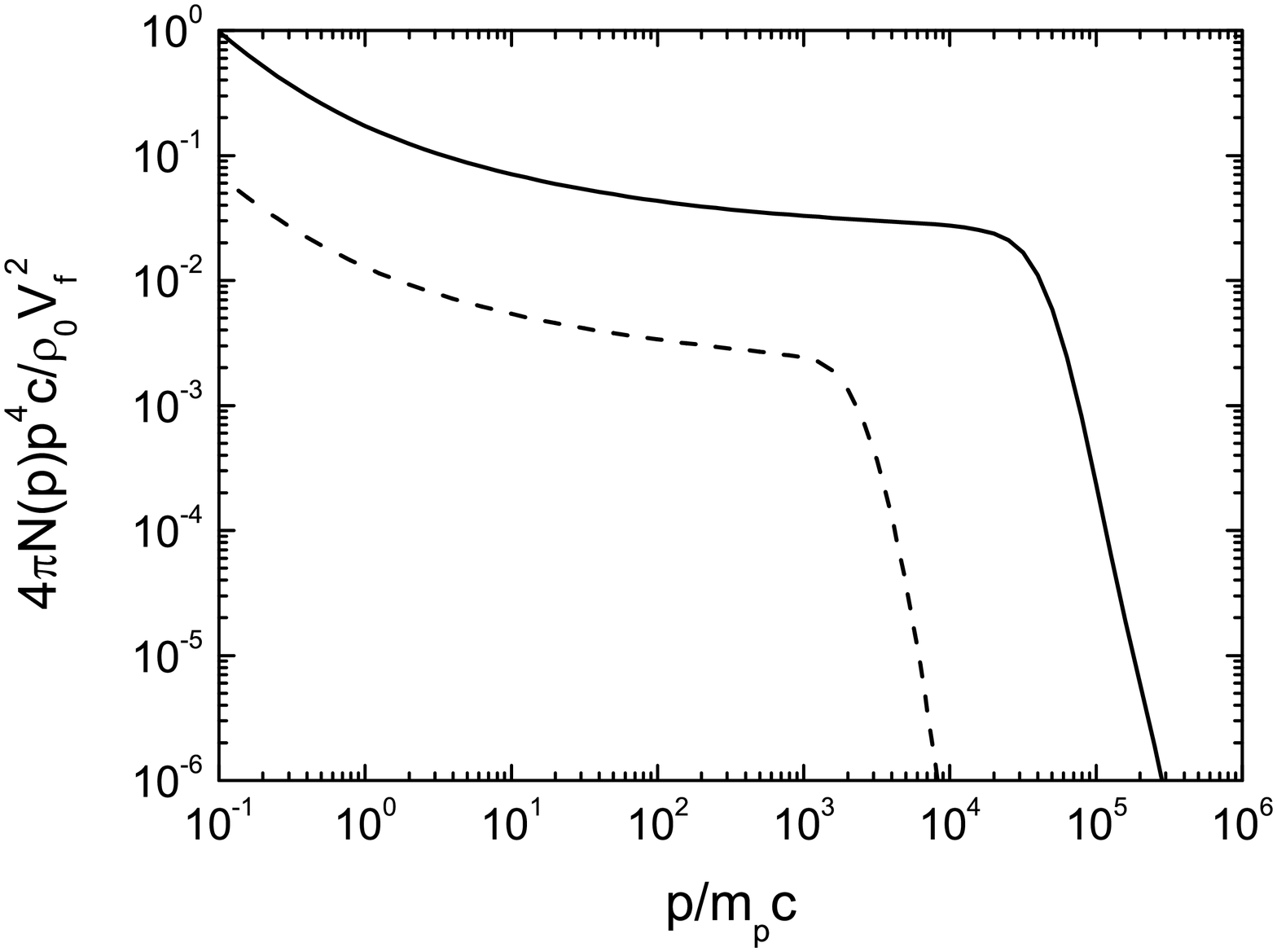}
\caption{Top panel: radial dependencies of the gas pressure (solid line), the CR pressure (dashed line), and the magnetic pressure (dotted line) at the epoch $t=2000$ yr,  RS, FS and CD are the positions of the forward shock, reverse shock and contact discontinue respectively.
Bottom panel: spectra of accelerated protons (solid line) and electrons (dashed line)in the position of forward shock at the epoch $t=2000$ yr.
}
\label{fig:1}
\end{figure}

\begin{figure}
  \begin{center}
\hspace{-0.90cm}
     \includegraphics[width=85mm,height=57mm]{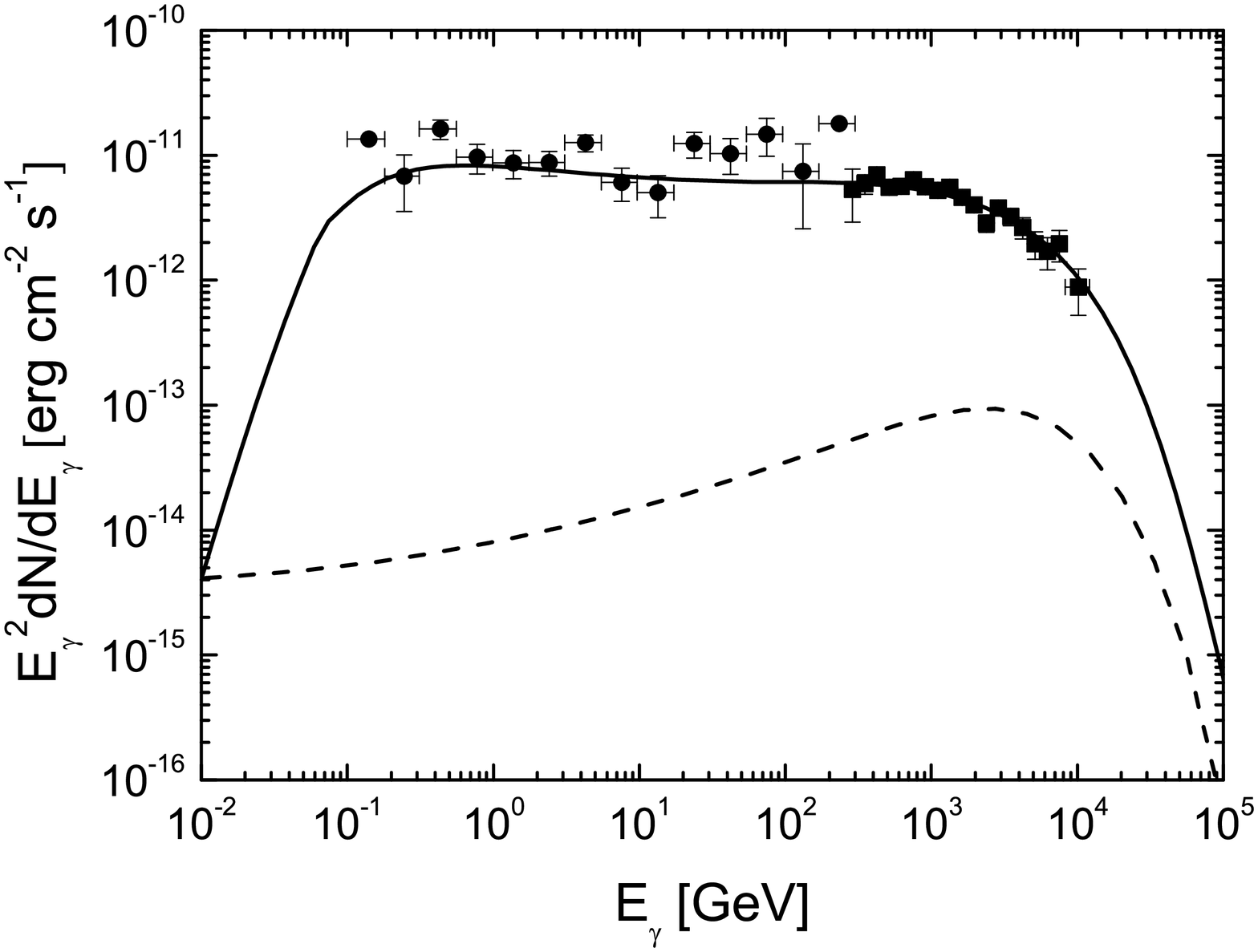}
  \end{center}
\caption{The high energy emission spectra of HESS J1640-465. Observed data of HESS J1640-465 derived from Abramowski et al. (2014a). The solid line and dashed line represent the very high energy emission from hadronic $\pi^0$ decay and leptonic IC scattering respectively.}
\label{fig:2}
\end{figure}

\subsection{HESS J1640-465}

HESS J1640-465 is a VHE-ray source observed by H.E.S.S., which is positionally coincident with the SNR G338.3-0.0 and evidently is a shell-type SNR with $8^\prime$ diameter \citep[e.g.][]{Aharonian2006,Whiteoak1996}. Based on 21 cm ${\rm H_I}$ absorption spectra to G338.3-0.0 and adjacent ${\rm H_{II}}$ regions, the distance to G338.3-0.0 is 12-13.5 kpc \citep{Lemiere2009}. While the distance is limited to a small range well, there is no direct measurement of the ambient density, but similar to RX J1713.7-3946, the HESS J1640-465 may evolve in a  wind bubble of progenitor star at present epoch, however some evidences for thermal radio emission in this SNR shell also indicate the presence of dense material \citep{castelletti11}. The age of SNR G338.3-0.0 is still ambiguous, and would be (5-8)kyr \citep{Slane2010}, recently \citet{Abramowski2014a} re-estimate the age would be (1-2) kyr which is significantly younger than the estimate of (5-8) kyr. Here we adopt an age of 2000 yr, a distance of 13 kpc, and an average number density of $n_0=0.2$ cm$^{-3}$.

In our calculations, the supernova explosion energy $E_{\rm sn}= 10^{51}$ erg and the initial ejecta mass $M_{\rm ej}= 1.4 M_{\odot}$ are assumed, the SNR shock propagates in the ambient medium with an average hydrogen number density $n_0=0.2$ cm$^{-3}$, a magnetic field strength $B_0=1.0~{\mu}G$, a temperature $T=10^{4}$ K, and the index of ejecta velocity distribution $k=7$. Below we use the injection parameter $\eta_{f}=\eta_{b}=5\times{10}^{-3}$ for protons, and the electron injection with $\eta_{f}=\eta_{b}={10}^{-3}$. The value of other parameters have been described in above section. The following results is obtained at the epoch $t=2000$ yr.

We present the radial dependencies of the gas pressure, the accelerated particle pressure, and the magnetic pressure in the top panel of Fig. \ref{fig:1}, where the accelerated particle, gas, and magnetic pressures at forward shock are $30\%$, $43\%$, and $8\%$ of the ram pressure. In the same figure we also show the forward shock (FS), reverse shock (RS) and contact discontinue (CD) positions, where the forward shock position $R_f=15.5$ pc is at the upper boundary of $4^\prime$ (about 15 pc) of radius for its observed value. The spectra of accelerated protons and electrons at the position of forward shock are shown in the bottom panel of Fig. \ref{fig:1}, the protons will be accelerated to run up to maximum energy $\sim 50$ TeV, which appear to be a source as significant contribution of Galactic cosmic ray flux around the knee. Because of the synchrotron losses of accelerated electrons in strong magnetic field, the maximum energy of electrons were only accelerated up to several TeV.

With above results, we calculate the spectra of high energy $\gamma$-rays which are produced by hadronic $\pi^0$ decay and leptonic IC scattering respectively. Our results are shown in Fig. \ref{fig:2}. The solid line represent the emission produced through the accelerated protons from G338.3+0.0 interacting with the ambient gas, and IC gamma-rays (dashed line) have been generated by accelerated electrons scattering off soft photons from 2.7 K microwave background radiation. For comparison, the observed spectrum of HESS J1640-465 is plotted in this figure. It can be seen that the observed GeV-TeV spectrum can be reproduced well in this model, and has a hadronic origin. Note that the leptonic contribution is not clear due to poorly X-ray observation.

\subsection{ HESS J1641-463}

\begin{figure}
\centering
\includegraphics[width=9.5cm, angle=0]{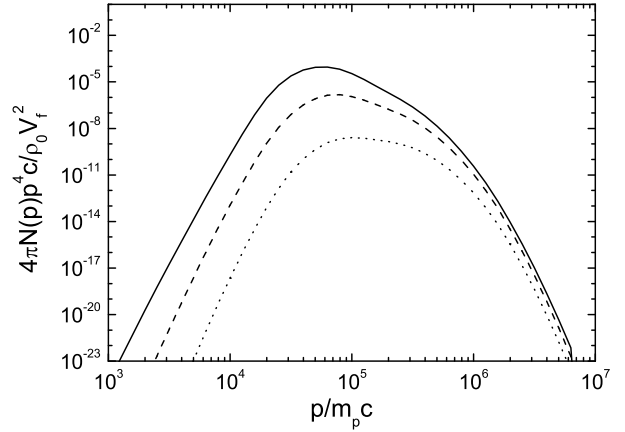}
\caption{Escaping particle spectra with $\delta=0.5$ from SNR are illustrated at the distances of $r=50$ (solid line), 100 (dashed line), and 200 (dotted line) pc for a given time $t=2000$ yr }
\label{fig:3}
\end{figure}

The protons which are accelerated up to high energy in SNR G338.3+0.0 can escape from the accelerated site into Galactic diffusion region. We calculate the escaped proton spectra for different diffusion distances at a given time $t=2000$ yr. Our results are shown in Fig.\ref{fig:3}, it can be seen that the energy flux of accelerated particles is sensitive to the diffusion distance, particular in the lower energy part of the spectrum. HESS J1641-463 is located only $0^\circ .25$ ( $r\sim 50$ pc for 13 kpc of a distance to observers) away from HESS J1640-465 \citep{Lemoine-Goumard2014}, below we used the particle spectra at the diffusion distance  $r=50$ pc to calculate the radiations from HESS J1641-463.

\begin{figure*}
  \begin{center}
  \begin{tabular}{cc}
\hspace{-0.90cm}
     \includegraphics[width=85mm,height=57mm]{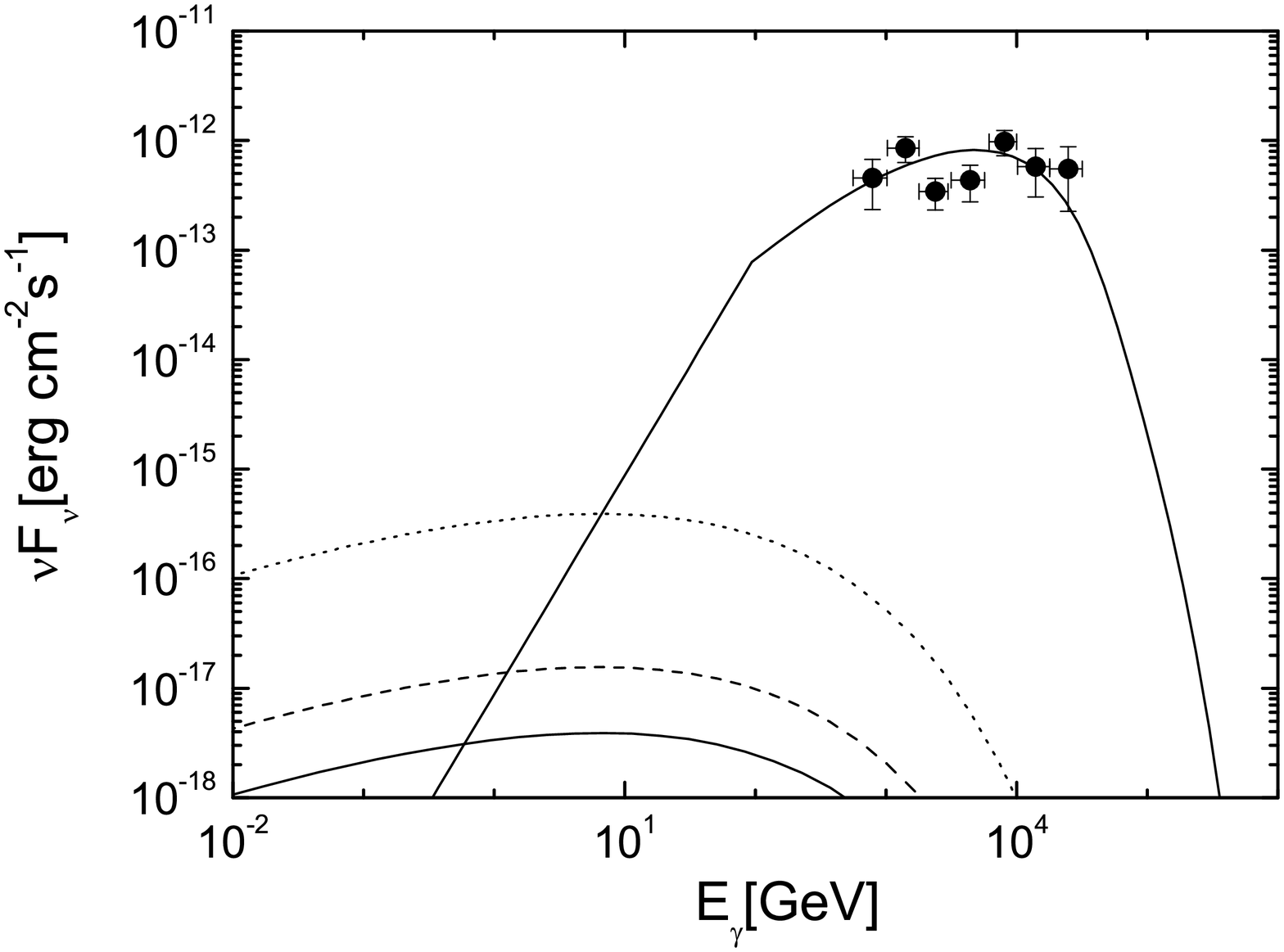} &
\hspace{-0.90cm}
     \includegraphics[width=85mm,height=57mm]{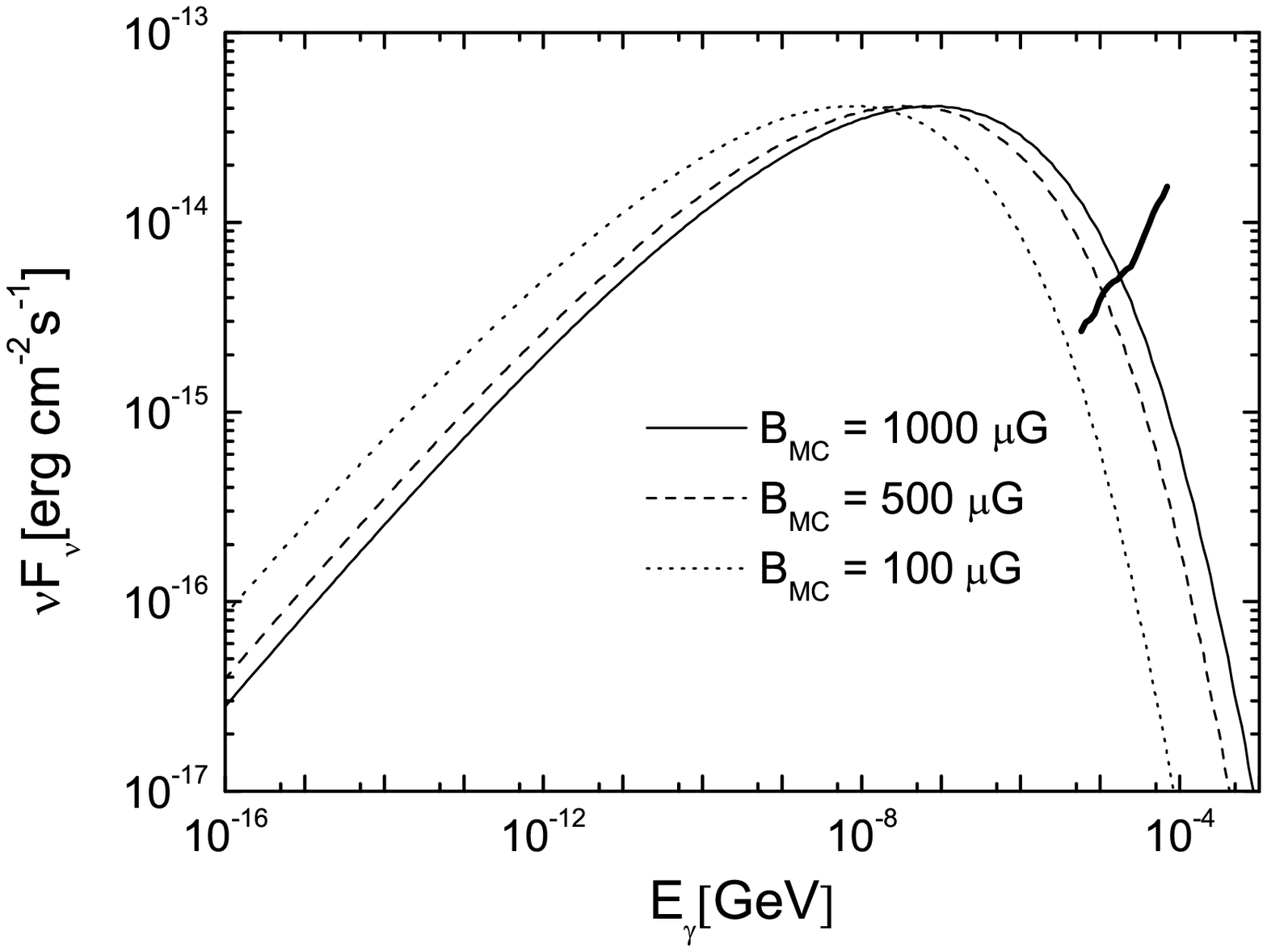}
\end{tabular}
  \end{center}
\caption{ SEDs of HESS J1641-463. Left panel: The high energy emission produced through the escaped energetic protons interacting with MC (thick solid line), and the secondary electron IC scattering in different magnetic field strength (their values are as same as those shown in right panel), observed data of HESS J1641-463 derived from Abramowski et al. (2014b). Right panel: possible secondary electron synchrotron emission for different magnetic field strength in MC, the thick solid line represents the sensibility of NuSTAR.}
\label{fig:4}
\end{figure*}

If these escaped protons go into a MC nearby the SNR, then they will interact with the the MC's medium. For a typical MC, its number density and mass of are $n_{\rm MC} = 100$ cm$^{-3}$ and $M_{\rm MC} = 10^5M_\odot$ respectively, we calculate the spectra of nonthermal photons which are produced by $\pi^0$ decay and inverse Compton scattering of the secondary electrons and positrons and show the results with the observed spectrum of HESS J1641-463 in the left panel of Fig.\ref{fig:4}. In this case, the gamma-ray emission mainly comes from $\pi^0$ decay in the proton-proton interaction (thick solid line), and inverse Compton scattering of the secondary $e^\pm$s also has a contribution which may dominate over lower energy region of gamma-rays and depends on the magnetic field strength (thin lines). However, our predicted flux of high energy gamma rays in Fermi/LAT energy range is lower than $10^{-13}~\rm erg~ cm^{-2}~ s^{-1}$ in the $\sim 100$ GeV band, a factor $10^3$ lower than the flux of 1FHL J1640.5-4634 at those energy band, even we plus the contribution from the secondary leptonic IC emission. A detection of GeV excess on the position of HESS J1641-463 would indicate either a contribution from an unrelated source or from a different component of radiation of the same source.

To explore the possible X-ray radiation from the source of HESS J1641-463, it is assumed that the X-rays come from synchrotron emission of secondary electron/positron pairs. The results are shown in the right of Fig.\ref{fig:4}. Obviously, the synchrotron emission flux level seems featureless, the only result is that the peak frequency is moving to high frequency with enlarging magnetic field strength. Particularly, if the magnetic field larger than $1000~\mu$Gs in the MC, then the peak energy will locate at a few keV region, and the flux of X-rays may be detected by NuSTAR.

\section{Summary and Discussion}

In this paper, we have investigated the TeV emissions from two TeV gamma-ray sources (HESS J1640-465 and HESS J1641-463) and given a self-consistent explanation in the modified non-linear diffusion shock acceleration model. On the one hand, the TeV emission from HESS J1640-465 is naturally produced through $\pi^0$ decay in the proton - proton interaction inside the SNR G338.3+0.0 ($r \le 2R_f$) and the inverse Compton scattering of the accelerated electrons inside the SNR has a negligible contribution (see Fig. \ref{fig:2}). On the other hand, high energy protons escaped from the SNR G338.3+0.0 undergo the Galactic diffusion and collide with the MC which is positionally coincident with HESS J1641-463, so the TeV emission of the HESS J1641-463 could come from the runaway protons interacting the MC at the region of $r > 2R_f$, the possible X-ray emission of the HESS J1641-463 is predicted as being relevant to secondary products of the runaway protons, it is worth looking forward that future detailed observations will provide important information for transparent explanation.

Finally, we would like to point out that we have given a self-consistent explanation of the TeV emissions from two TeV gamma-ray sources in the scenario which is one of the scenarios proposed by \citet{Abramowski2014b}, but other possible scenarios are not excluded.

\acknowledgments
\section*{acknowledgments}
 We would like to thank the anonymous referee for his/her very instructive comments. This work is partially supported by the National Science Foundation of China (11303012, 11433004, 11133006, 11163006, 11173054), the Natural Science Foundation of Yunnan province of
China (2013FZ100), the Strategic Priority Research Program, and the Emergence of Cosmological Structures of the Chinese Academy of Sciences, Grant No. XDB09000000.

\clearpage

\end{document}